\documentstyle[floats,preprint,prb,psfig,aps,tighten]{revtex}
\parskip=\medskipamount
\begin{document}
\input{epsf} 
\draft
\renewcommand{\textfraction}{0.0}
\renewcommand{\dblfloatpagefraction}{0.8}
\renewcommand{\topfraction}{1.0}   
\renewcommand{\bottomfraction}{1.0}   
\renewcommand{\floatpagefraction}{0.8}
\def\ep{\epsilon}
\def\al{\alpha}
\def\be{\beta}
\def\ga{\gamma}
\def\la{\lambda}
\def\th{\theta}
\def\de{\delta}
\def\si{\sigma}
\def\ti{\tilde}
\def\et{\eta}
\def\pa{\partial}
\def\om{\omega}
\def\fr{\frac}
\def\be{\begin{equation}}
\def\ee{\end{equation}}
\def\bea{\begin{eqnarray}}
\def\eea{\end{eqnarray}}
\title{Asymmetric gap soliton modes in diatomic lattices\\ 
       with cubic and quartic nonlinearity}
\author{Guoxiang Huang$^{1,2,3}$ and Bambi Hu$^{2,4}$ }
\address{ $^{1}$Max-Planck-Institut f\"{u}r Physik komplexer Systeme,
         N\"{o}thnitzer Strasse 38,\\ D-01187 Dresden, Germany\\
          $^{2}$Center for Nonlinear Studies and Department of Physics, Hong Kong
         Baptist University,\\ Hong Kong, China\\        
          $^{3}$Department of Physics, East China Normal University,
         Shanghai 200062,
         China\\
          $^{4}$Department of Physics, University of Houston, Houston TX 77204,
         USA}
\maketitle
\begin{abstract}
Nonlinear localized excitations in one-dimensional diatomic lattices with 
cubic and quartic nonlinearity are considered analytically by a 
quasi-discreteness approach. The criteria for the occurence of
asymmetric gap solitons\,(with vibrating frequency lying in the gap of 
phonon bands) and small-amplitude, asymmetric
intrinsic localized modes\,(with the 
vibrating frequency being
above all the phonon bands) are obtained explicitly based on the
modulational instabilities of corresponding linear lattice plane waves.
The expressions of particle  displacement for all these
nonlinear localized excitations are also given. The result is applied to
standard two-body potentials of the Toda, Born-Mayer-Coulomb, 
Lennard-Jones, and Morse type. The comparison with previous
numerical study of the anharmonic gap modes in diatomic lattices for 
the standard two-body potentials is made and good agreement is found.
\end{abstract}
\section{INTRODUCTION}
The study of the dynamics of nonlinear lattices and related solitonic
excitations has greatly influenced by the pioneering works of Fermi,
Pasta, and Ulam,\cite{fer} and of Zabusky and Kruskal. \cite{zab1}  Most
of the work in this area has focused on models of one-dimensional\,(1D)
monatomic chains with simple interatomic potentials of polynomial 
\cite{zab2,col,pne1,fly1,fly2}, which can approximate any realistic 
potential near the equilibrium separation distance of two atoms. 
This description, usually done in a  continuum limit, is only valid
for a zone-boundary phonon mode, {\it i.\,e}. for $q$, the wavenumber
of lattice waves, being near zero or $\pi/d_0$, where $d_0$ is lattice
spacing. 
In 1972, Tsurui\cite{tsu} 
proposed an analytical method for studying the nonlinear excitations
of lattices valid in the whole Brillouin zone\,(BZ). Later on this
approach  was extended by Remoissenet\cite{rem} and Huang.\cite{hua1,hua2}
Exact analytical solutions for the nonlinear localized excitations in 1D 
monatomic lattices can be obtained only for the Toda\cite{tod} and 
Ablowitz and Ladik\cite{abl} lattices, which are discrete completely
integrable systems.

In recent years, the interest in localized excitations in nonlinear lattices
has been renewed  due to the identification of a new type of 
anharmonic 
localized modes.\cite{hua1,dol,sie1,pag,bur,kiv1,san,fla1,kis1,sie2,fla2}
These modes, called the intrinsic localized modes\,(ILM's),\cite{sie1} or the
discrete breathers,\cite{fla2} are the discrete analog of the envelope\,(or
breather) solitons with their spatial extension being only of a few lattice 
spacing and the vibrating frequency lying above the upper cutoff
of phonon bands.\cite{yos}
Experimentally, the ILM's have been observed in coupled pendulum 
lattices\cite{che1} and electrical lattices.\cite{mar} The quantum 
mechanical aspects of the ILM's have also been considered.\cite{shi,hua3,ros}
However, examination of the 1D lattices with standard Toda, Born-Mayer-Coulumb,
Lennard-Jones, and Morse two-body interatomic potentials demonstrates that
the ILM's do not appear above the top of the plane wave spectrum. 
The physical reason for this is that the cubic nonlinearity in the Taylor
expansion of these realistic potentials is too strong. One of the effects
of the cubic nonlinearity is that increasing the magnitude of the cubic
term makes the potentials softer and hence decreases the localized
mode frequency. The localized mode is destroyed as it approaches
the bounding plane wave spectrum.\cite{hua1,bic}

Recently, much attention has been paid to the gap solitons in nonlinear
diatomic
lattices.\cite{hua2,shi,kiv2,chu,kis2,kis3,fra,aok,bon,tei,kon,hua4,kis4}
The concept of the gap solitons was first introduced by Chen and 
Mills\cite{che2} when investigating the nonlinear optical response of
superlattices. For a diatomic lattice, the phonon spectrum of the system
consists of two branches\,(acoustic and optical ones), induced by mass
or force-constant difference of two kinds of particles. Due to nonlinearity
gap soliton modes  may appear as localized excitations with 
vibrating frequency 
being in the gap of the linear spectrum. Since the gap solitons occur in
perfect lattices with discrete translational symmetry, a name ``anharmonic
gap mode'' or ``intrinsic gap mode\,(IGM)'' was given by Sievers and his
collaborators.\cite{kis2,kis3,kis4} It is possible that the ILM's and the 
IGM's may be created experimentally in diatomic 
lattices. Refs.\,43 and 44 reported
some experimental studies of the gap solitons, resonant kinks and the ILM's
in a damped and parametrically excited 1D diatomic pendulum lattices.

Since the standard two-body potentials of the Toda, Born-Mayer-Coulomb,
Lennard-Jones, and Morse type have a strong cubic nonlinearity in their
Taylor expansion near equilibrium position, it is therefore
 necessary to consider
the nonlinear excitations in the diatomic lattices with cubic and
quartic anharmoniticity. In their recent contributions, Kiselev
{\it et al}.\cite{kis2,kis3} investigated the anharmonic localized modes 
in  1D diatomic lattices with above mentioned two-body  potentials.
By using a rotating wave approximation combined with computer simulation,
they showed that an ILM does {\it not} exist and a nonlinear optical lower
cutoff gap mode\,({\it i.\,e}. IGM) is general feature of these diatomic 
lattices. Lately, Franchini {\it et al}.\cite{fra} numerically found
that for a potential with the cubic and quartic nonlinearity there exists
a ``critical'' $K_3$\,(cubic force constant in the potential) value.
For small $K_3$, nonlinear optical and acoustic upper cutoff
localized modes occur, while for large $K_3$ these modes disappear 
and a nonlinear optical lower cutoff mode rises. 
In a recent work, Bonart {\it et al}.\cite{bon1} investigated the boundary 
condition effects in the diatomic lattices with cubic and quartic 
anharmonicity. Based on a rotating wave approximation they gave 
existence criteria for the ILM's and IGM's, which are related to the
stability properties of linear optical upper and lower cutoff phonon modes.
These studies
posed an interesting problem of how to provide an analytical approach
which can give not only the explicit criteria for the existence of 
the ILM's and the
IGM's as well as other possible nonlinear eacitations for both optical
and acoustic branches but also the approximate analytical
expressions for these nonlinear excitations in a unified way. 
It is this problem that will be addressed here.

There are several theoretical methods to study the nonlinear localized
excitations in diatomic lattices\,(see Ref.\,11 and references therein).
In this paper we use the quasi-discreteness approach\,(QDA) for diatomic
lattices\cite{hua2} to investigate the ILM's and IGM's as well as
kink-like excitations with small-amplitude 
in 1D diatomic lattices with cubic and quartic
nonlinear interactions between their nearest- neighbor particles. 
The paper is organized as follows. In Sec.\,II, the model is introduced
and an asymptotic expansion based on the QDA is made for the equations
of motion. By using the results obtained in Sec.\,II, in Sec.\,III we 
discuss the solutions of the 
ILM's and IGM's  in a simple and unified way. Some explicit
criteria and expressions of particle displacement of the ILM's and the
IGM's are also given in this section. In Sec.\,IV we apply our results
to the standard two-body potentials from Toda to Morse type and make a 
comparison with existed  numerical experiments. Finally, Sec.\,V 
contains a discussion and summary of our results.
%
%
%
\section{MODEL AND ASYMPTOTIC EXPANSION}
\subsection{The model}
We consider a 1D diatomic lattices with a nearest-neighbor interaction
between particles. The restriction to the nearest-neighbor interaction
is for simplicity and the approach can be easily extended to second 
and higher neighbors.  
The Hamiltonian of the system is given by
\be
H=\sum_i\left[ \fr{1}{2}m_{i}\left (\fr{{\rm d}u_i}{{\rm d}t}\right )^2
  +V(u_{i+1}-u_i) \right],
\ee
where $u_i=u_i(t)$ is the displacement from its equilibrium position
of the $i$th particle with mass $m_i
=m\delta_{i,2k}+M\delta_{i,2k+1}\,(M>m, k$ is an integer). 
The potential $V(r)$ is quite general, typically it can be
the standard two-body potentials of the Toda, Born-Mayer-Coulomb, 
Lennard-Jones and Morse type\,(for their detailed expressions, see 
Sec.\,IV below). We focus on displacements with smaller amplitude
which can be detected experimentally without introducing
reconstruction or phase transitions in the system. This allows
us to Tayler expand the potential $V(r)$ at the  equilibrium position $r=0$ 
in a power series of the displacements to four order.\cite{fra} Thus we 
obtain an approximate $K_2$-$K_3$-$K_4$ potential
\be
V(r)=\fr{1}{2}K_2r^2+\fr{1}{3}K_3 r^3+\fr{1}{4}K_4 r^4,
\ee
where $K_2(>0)$, $K_3$ and $K_4(>0)$ are harmonic, cubic and quartic force
constants, respectively. We assume that the basic features of the
weakly nonlinear localized excitations for the standard two-body potentials
may  be obtained by corresponding the $K_2$-$K_3$-$K_4$ potentials. Then
the Hamiltonian (1) takes the following form
\be
H=\sum_{i}\left [\fr{1}{2}m_{i}\left (\fr{{\rm d}u_i}{{\rm d}t}\right )^2
+\fr{1}{2}K_2(u_{i+1}-u_i)^2
+\fr{1}{3}K_3(u_{i+1}-u_i)^3
+\fr{1}{4}K_4(u_{i+1}-u_i)^4\right ].
\ee
Since each of the standard two-body potentials mentioned above  
has only one minimum, we assume that for the $K_2$-$K_3$-$K_4$ potential
(2) there is the constraint
\be
\fr{K_{3}^{2}}{K_2K_4}<4,
\ee
unless there are two minima\,({\it i.\,e}. double well potential) hence
the system may admit some types of nonlinear excitations which will
not be discussed here.

If we write $u_{2k}=v_n$\,(even particles) and 
$u_{2k+1}=w_n$\,(odd particles), $n$ is the index of the $n$th unit cell
with a lattice spacing $d=2d_0$, $d_0$ is the equilibrium distance
between two adjacent particles, the system can be split into two sublattices.
The equations of motion for $v_n$ and $w_n$ are
\bea
&  &  m\fr{{\rm d}^2}{{\rm d}t^2}v_n=
      K_2(w_n+w_{n-1}-2v_n)
      +K_3[\,(w_n-v_n)^2-(w_{n-1}-v_n)^2]\nonumber\\
&  &  \hspace{19mm}+K_4[\,(w_n-v_n)^3+(w_{n-1}-v_n)^3],\\
&  &  M\fr{{\rm d}^2}{{\rm d}t^2}w_n=
      K_2(v_n+v_{n+1}-2w_n)
      -K_3[\,(v_n-w_n)^2-(v_{n+1}-w_n)^2]\nonumber\\
&  &  \hspace{20mm}+K_4[\,(v_n-w_n)^3+(v_{n+1}-w_n)^3].
\eea
The linear dispersion relation of Eqs.\,(5) and (6) is
\be
\omega_{\pm}(q)=\{I_2+J_2\pm[(I_2+J_2)^2
-4I_2J_2\sin^2(qd/2)]^{\frac{1}{2}}\}^{\frac{1}{2}},
\ee
where $I_2=K_2/m$ and $J_2=K_2/M$. The minus\,(plus) sign corresponds to 
acoustic\,(optical) mode. At wave number $q=0$ the eigen frequency spectrum 
has a lower cutoff $\om_{-}(0)=0$ for the acoustic mode and an upper cutoff
$\om_{+}(0)\equiv \om_{3}=[2(I_2+J_2)]^{1/2}$ for the optical mode. At 
$q=\pi/d$ there exists a frequency gap between the upper cutoff
of the acoustic branch, $\om_{-}(\pi/d)\equiv \om_1=\sqrt{2J_2}$, and the lower
cutoff of the optical branch, $\om_{+}(\pi/a)\equiv \sqrt{2I_2}$. The width 
of the frequency gap is 
$\sqrt{2I_2}-\sqrt{2J_2}=\sqrt{2K_2}(1/\sqrt{m}-1/\sqrt{M})$.
In linear theory, the amplitudes of lattice waves are constants and  linear
waves can not propagate and will be damped when $\om$\,(the frequency of the
waves) lies in the regions $\om_1<\om<\om_2$ and $\om>\om_3$. Accordingly, these
regions are the ``forbidden bands'' of the linear waves. This property of
the eigen frequency spectrum results from the discreteness of the 
system\,({\it i.\,e}. discrete translational symmery). 
However, when the nonlinearity in Eqs.\,(5) and (6) is considered, the above
conclusions are no longer valid. As localized excitations, some nonlinear
modes may appear, whose oscillatory frequencies can lie in these forbidden
bands of the phonon spectrum.
\subsection{Asymptotic expansion}
We use the QDA for diatomic lattices  developed in Ref.\,11 to investigate the
effects of nonlinearity and discreteness of the system. In this treatment one 
sets
\be
u_n(t)=\ep u^{(1)}_{n,n}+\ep^2 u^{(2)}_{n,n}
             +\ep^3 u^{(3)}_{n,n}+\cdots\nonumber\\
\ee
where $\ep$ is a  smallness and ordering parameter denoting the relative 
amplitude of the excitation and
$u_{n,n}^{(\nu)}=u^{(\nu)}(\xi_n,\tau;\phi_n)$. 
$\xi_n=º\ep (na-\la t)$ and $\tau=\ep^2t$ 
are two multiple-scales variables\,(slow variables). 
$\la$ is a
parameter to be determined by a solvability condition. The ``fast'' variable, 
$\phi_n=qnd-\om (q)t$, representing the phase of carrier wave, is taken to be 
completely discrete.  Substituting (8) into (5) and  (6) and comparing the
power of $\ep$, we obtain a hierarchy of equations about $v_{n,n}^{(j)}$ and 
$w_{n,n}^{(j)}$\,($j=1,\,2,\,3,\cdots$):
\bea
&  &  \fr{\pa^2}
      {\pa t^2}v_{n,n}^{(j)}-I_2(w_{n,n}^{(j)}+w_{n,n-1}^{(j)}-2v_{n,n}^{(j)})
      =M_{n,n}^{(j)}
\eea
with
\bea
&  &  M_{n,n}^{(1)}=0,\\
&  &  M_{n,n}^{(2)}=2\la \fr{\pa^2}{\pa t\pa \xi_n}v_{n,n}^{(1)}-I_2d\fr{\pa}{\pa   
      \xi_n}w_{n,n-1}^{(1)}+I_3(w_{n,n}^{(1)}-v_{n,n}^{(1)})^2
      -I_3(w_{n,n-1}^{(1)}-v_{n,n}^{(1)})^2,\\
&  &  M_{n,n}^{(3)}=2\la \fr{\pa^2}{\pa t\pa \xi_n}v_{n,n}^{(2)}
      -\left (2\fr{\pa^2}{\pa t\pa\tau}+\la^2 \fr{\pa^2}{\pa\xi_{n}^2}\right )
      v_{n,n}^{(1)}+I_2 \left ( -d\fr{\pa}{\pa \xi_n} w_{n,n-1}^{(2)}
      +\fr{d^2}{2!}\fr{\pa^2}{\pa \xi_{n}^2}w_{n,n-1}^{(1)}\right )
      \nonumber\\
&  &  \hspace{1.5cm}+2I_3(w_{n,n}^{(1)}-v_{n,n}^{(1)}) (w_{n,n}^{(2)}-v_{n,n}^{(2)})
      -2I_3 (w_{n,n-1}^{(1)}-v_{n,n}^{(1)}) (w^{(2)}_{n,n-1}-v^{(2)}_{n,n}
      -d\fr{\pa}{\pa\xi_n}w^{(1)}_{n,n-1})\nonumber\\
&   &  \hspace{1.5cm}+I_4 [ (w^{(1)}_{n,n}-v^{(1)}_{n,n})^3 
       + (w^{(1)}_{n,n-1}-v^{(1)}_{n,n})^3],\\
&  &  \vdots\vdots \nonumber
\eea
and
\bea
&  &  \fr{\pa^2}{\pa t^2}w_{n,n}^{(j)}-J_2(v_{n,n}^{(j)}+v_{n,n+1}^{(j)}
      -2w_{n,n}^{(j)})
      =N_{n,n}^{(j)}
\eea
with
\bea
&  &  N_{n,n}^{(1)}=0,\\
&  &  N_{n,n}^{(2)}=2\la \fr{\pa^2}{\pa t\pa \xi_n}w_{n,n}^{(1)}+J_2d\fr{\pa}{\pa   
      \xi_n}v_{n,n+1}^{(1)}-J_3(v_{n,n}^{(1)}-w_{n,n}^{(1)})^2
      +J_3(v_{n,n+1}^{(1)}-w_{n,n}^{(1)})^2,\\
&  &  N_{n,n}^{(3)}=2\la \fr{\pa^2}{\pa t\pa \xi_n}w_{n,n}^{(2)}
      -\left (2\fr{\pa^2}{\pa t\pa\tau}+\la^2 \fr{\pa^2}{\pa\xi_{n}^2}\right )
      w_{n,n}^{(1)}+J_2 \left ( d\fr{\pa}{\pa \xi_n} v_{n,n+1}^{(2)}
      +\fr{d^2}{2!}\fr{\pa^2}{\pa \xi_{n}^2}v_{n,n+1}^{(1)}\right )
      \nonumber\\
&  &  \hspace{1.5cm}-2J_3(v_{n,n}^{(1)}-w_{n,n}^{(1)}) (v_{n,n}^{(2)}-w_{n,n}^{(2)})
      +2J_3 (v_{n,n+1}^{(1)}-w_{n,n}^{(1)}) (v^{(2)}_{n,n+1}-w^{(2)}_{n,n}
      +d\fr{\pa}{\pa\xi_n}v^{(1)}_{n,n+1})\nonumber\\
&   &  \hspace{1.5cm}+J_4 [ (v^{(1)}_{n,n}-w^{(1)}_{n,n})^3 
       + (v^{(1)}_{n,n+1}-w^{(1)}_{n,n})^3],\\
&  &  \vdots\vdots \nonumber
\eea
which can be solved order by order. In Eqs.\,(9)-(16), we have
defined $I_{i}=K_{i}/m$
and $J_{i}=K_{i}/M$\,($i=2,\,3,\,4$). The expressions of 
$M_{n,n}^{(j)}$ and $N_{n,n}^{(j)}$\,($j=4,\,5,\cdots$) need not be 
written down explicitly here.
\subsection{Amplitude equations for acoustic and optical modes}
In order to avoid possible divergence for zone-boundary
phonon modes, we solve the acoustic and optical modes
separetely. First we consider the low-frequency acoustic mode of the
system. For this we rewrite Eqs.\,(9) and (13) in the form
\bea
&   &  \hat{L}w_{n,n}^{(j)}=J_2( M_{n,n}^{(j)}+M_{n,n+1}^{(j)} )
       +\left (\fr{\pa^2}{\pa t^2}+2I_2\right )N_{n,n}^{(j)},\\
&   &  \left (\fr{\pa^2}{\pa t^2}+2I_2\right )v_{n,n}^{(j)}=I_2
       \left( w_{n,n}^{(j)}+w_{n,n-1}^{(j)}\right )+M_{n,n}^{(j)},
\eea
where the operator $\hat{L}$ is defined by
\be
\hat{L}u_{n,n}^{(j)}=\left(\fr{\pa^2}{\pa t^2}+2I_2\right )
                     \left(\fr{\pa^2}{\pa t^2}+2J_2\right )
                     u_{n,n}^{(j)}-I_2J_2\left( u_{n,n-1}^{(j)}+
                     u_{n, n+1}^{(j)}+2u_{n,n}^{(j)}\right)
\ee
with $u_{n,n}^{(j)}(j=1,\,2,\,3,\cdots)$ a  set of arbitrary functions.
For $j=1$ it is easy to get
\bea
&  &  w_{n,n}^{(1)}=F_{10}(\tau,\xi_n)+[\,F_{11}(\tau,\xi_n){\rm e}^{i\phi_{n}^{-}}
      +{\rm c.\,c.}],\\
&  &  v_{n,n}^{(1)}=F_{10}(\tau,\xi_n)
      -\left[\fr{I_2(1+{\rm e}^{-iqd})}{\om_{-}^{2}-2I_2}F_{11}(\tau,\xi_n)
      {\rm e}^{i\phi_{n}^{-}}
      +{\rm c.\,c.}\right ]
\eea
with $\phi_n^{-}=qnd-\om_{-}(q)t$. $\om_{-}(q)$ has been given in Eq.\,(7)
with a minus sign. The amplitude\,(or envelope) functions $F_{10}$ and
$F_{11}$ are yet to be determined. $F_{10}$ is a real function representing
the ``direct current\,(dc)'' part relative to the fast variable $\phi_n^{-}$
and $F_{11}$ is a  complex amplitude of the ``alternating current\,(ac)''
part. If $K_3=0$, the dc part\,($F_{10}$) vanishes in this order. For
$j=2$\,(the second order) a solvability condition determines 
$\la =V_{g}^{-}={\rm d}\om_{-}/{\rm d}q$\,({\it i.\,e}. the group velocity
of the carrier waves) thus $\xi_n=\xi_n^{-}\equiv \ep (nd-V_{g}^{-}t)$.  
In the third order\,($j=3$), solvability conditions 
 yields the  evolution equations for $F_{10}$ and $F_{11}$:
\bea
&  &  i\fr{\pa}{\pa \tau}F_{11}+\fr{1}{2}\al_{-}\fr{\pa}{\pa \xi_{n}^{-}}
      \fr{\pa}{\pa \xi_{n}^{-}}F_{11}+\beta_{-}F_{11}\fr{\pa}{\pa \xi_{n}^{-}}F_{10}
      +\ga_{-}\,|F_{11}|^{2}F_{11}=0,\\
&  &  \de_{-}\fr{\pa}{\pa \xi_{n}^{-}}\fr{\pa}{\pa \xi_{n}^{-}}F_{10}
      +\si_{-}\fr{\pa}{\pa \xi_{n}^{-}}\,|F_{11}|^2=0.
\eea
The detailed expressions of the 
coefficients in  Eqs.\,(22) and
(23) are given in Appendix A.

Second, we study the high-frequency optical mode excitations. In this case we
recast Eqs.\,(9) and (13) into the form
\bea
&   &  \hat{L}v_{n,n}^{(j)}=J_2( N_{n,n}^{(j)}+N_{n,n-1}^{(j)} )
       +\left (\fr{\pa^2}{\pa t^2}+2J_2\right )M_{n,n}^{(j)},\\
&   &  \left (\fr{\pa^2}{\pa t^2}+2I_2\right )w_{n,n}^{(j)}=J_2
       \left( v_{n,n}^{(j)}+v_{n,n+1}^{(j)}\right )+N_{n,n}^{(j)}.
\eea
By the same procedure of solving the acoustic mode given above, we
obtain
\bea
&  &  v_{n,n}^{(1)}=G_{10}(\tau,\xi_n)+[\,G_{11}(\tau,\xi_n){\rm e}^{i\phi_{n}^{+}}
      +{\rm c.\,c.}],\\
&  &  w_{n,n}^{(1)}=G_{10}(\tau,\xi_n)
      -\left[\fr{J_2(1+{\rm e}^{iqd})}{\om_{+}^{2}-2J_2}G_{11}(\tau,\xi_n)
      {\rm e}^{i\phi_{n}^{+}}
      +{\rm c.\,c.}\right ]
\eea
with $\phi_n^{+}=qnd-\om_{+}(q)t$. The evolution equations for the amplitudes 
$G_{10}$\,(dc part of the optical mode) and $G_{11}$\,(complex amplitude
for ac part of the optical mode) are given by
\bea
&  &  i\fr{\pa}{\pa \tau}G_{11}+\fr{1}{2}\al_{+}\fr{\pa}{\pa \xi_{n}^{+}}
      \fr{\pa}{\pa \xi_{n}^{+}}G_{11}+\beta_{+}G_{11}\fr{\pa}{\pa \xi_{n}^{+}}G_{10}
      +\ga_{+}\,|G_{11}|^{2}G_{11}=0,\\
&  &  \de_{+}\fr{\pa}{\pa \xi_{n}^{+}}\fr{\pa}{\pa \xi_{n}^{+}}G_{10}
      +\si_{+}\fr{\pa}{\pa \xi_{n}^{+}}\,|G_{11}|^2=0,
\eea
where  $\xi_{n}^{+}=\ep (nd-V_{g}^{+}t)$ with 
$\la =V_{g}^{+}={\rm d}\om_{+}/{\rm d}q$. 
The coefficients in Eqs.\,(28) and
(29) are also given in the Appendix A.

Under the transformation
\bea
&  &  F_{10}=(1/\ep)g_{-},\hspace{1cm}F_{11}=(1/\ep)f_{-},\\
&  &  G_{10}=(1/\ep)g_{+},\hspace{1cm}G_{11}=(1/\ep)f_{+},
\eea
the nonlinear amplitude equations
(22), (23), (28) and (29) can be written in the unified form
\bea
&  &  i\fr{\pa}{\pa t}f_{\pm}+\fr{1}{2}\al_{\pm}\fr{\pa}{\pa x_{n}^{\pm}}
      \fr{\pa}{\pa x_{n}^{\pm}}f_{\pm}+\beta_{\pm}f_{\pm}\fr{\pa}
      {\pa x_{n}^{\pm}}g_{\pm}
      +\ga_{\pm}\,|f_{\pm}|^{2}f_{\pm}=0,\\
&  &  \de_{\pm}\fr{\pa}{\pa x_{n}^{\pm}}\fr{\pa}{\pa x_{n}^{\pm}}g_{\pm}
      +\si_{\pm}\fr{\pa}{\pa x_{n}^{\pm}}\,|f_{\pm}|^2=0,
\eea
when returning to the original variables. In Eqs.\,(32) and (33),
 $x_{n}^{\pm}=nd-V_{g}^{\pm}t$ and
plus\,(minus) sign corresponds to the optical\,(acoustic) mode, respectively.

Finally, we consider the acoustic mode at $q=0$. 
Noting that Eq.\,(32) for minus sign is invalid at $q=0$ for the description
of nonlinear excitations since $\beta_{-}|_{q=0}=\gamma_{-}|_{q=0}=0$
and $\al_{-}|_{q=0}=\infty$. This breakdown is due to the fact that at $q=0$
an  acoustic mode excitation is a long wavelength one.   
In this case a discrete long-wave approximation\cite{hua2} 
 should be applied.  By using the same technique used in 
Ref.\,11, for the acoutic mode at $q=0$ we obtain a long wavelength amplitude
equation
\be
\fr{\pa u}{\pa t}+P u\fr{\pa}{\pa x_{n}}u+Q u^{2}\fr{\pa}{\pa x_{n}}u
+H \fr{\pa^3}{\pa x_{n}^{3}}u=0,
\ee
where $u=\pa A_{0}/\pa x_{n}$, $x_{n}=nd-ct$ with $c^{2}=K_{2}d^{2}/[2(M+m)]$.
$A_{0}$ is the leading order approximation of $v_{n}$ and $w_{n}$. 
Since $\om_{-}(0)=0$, for the long wavelength acoustic mode there is
no carrier wave.  Thus the displacement of the lattice is purely
a ``direct current''. Eq.\,(34) without the second term $P\,\pa u/\pa x_n$
is standard modified Korteweg-de Vries\,(MKdV) equation. Thus (34)
is a {\it modified} MKdV\,(MMKdV) equation. Its 
coefficients are given by
\bea
&  &  P=\fr{d^2}{4}\fr{K_3}{K_2}\left (\fr{2K_2}{M+m}\right )^{\fr{1}{2}},\\
&  &  Q=
      \fr{3d^3}{16}\fr{K_4}{K_2}\left(\fr{2K_2}{M+m}\right)^{\fr{1}{2}},\\
&  &  H=\fr{d^3}{16}\left (\fr{2K_2}{M+m}\right )^{\fr{1}{2}}
      \left [\fr{1}{3}-\fr{mM}{(M+m)^2}\right ].
\eea
\section{ASYMMETRIC GAP SOLITONS, KINKS  AND 
INTRINSIC LOCALIZED MODES}
When deriving the nonlinear 
amplitude equations (32)-(34) we have not used any kind of 
decoupling ansatz  for the motion of two kinds of particles with 
different mass. This is one of the advantages of the QDA. On the other hand,
the nonlinear amplitude equations, which are the reduced forms of the original
equations of motion (5) and (6) for small-amplitude excitations, 
are valid in the whole BZ\,($-\pi/d<q\leq\pi/d$)
except a zero-dispersion point for the optical phonon branch.\cite{for} 
Thus one can obtain the gap solitons and ILM's as well as some possible
new nonlinear excitations by solving the cutoff modes of the system in
a simple and unified way.

\vspace{2mm}

(1). {\bf Optical upper cutoff mode}. For the optical mode at $q=0$, we 
have $\om_{+}=\om_3=[2K_2(1/m+1/M)]^{1/2}$, $V_{g}^{+}=0$, 
$x_{n}^{+}=nd\equiv x_n$,
$\al_{+}=-K_{2}d^{2}/[2(M+m)\om_{3}]$, $\beta_{+}=-K_{3}\om_{3}d/(2K_{2})$, 
$\ga_{+}=-3K_{4}\om_{3}(1+m/M)^2/(2K_2)$,
$\de_{+}=K_{2}^{2}d^{2}/(Mm)$, and $\si_{+}=4K_{2}K_{3}d(1+m/M)^2/(Mm)$.
Eqs.\,(32) and (33) with plus sign take the form
\bea
&  &  i\fr{\pa}{\pa t}\tilde{f}_{+}+\fr{1}{2}\al_{+}\fr{\pa^2}{\pa x_{n}^{2}}
      \tilde{f}_{+}
      +\tilde{\ga}_{+}\,|\tilde{f}_{+}|^{2}\tilde{f}_{+}=0,\\
&  &  \fr{\pa}{\pa x_{n}}g_{+}=
      -\fr{\si_{+}}{\de_{+}}|\tilde{f}_{+}|^2+C_1,
\eea
where $\ti{f}_{+}=f_{+}\exp (-i\beta_{+}C_{1}t)$ with $C_1$ 
an integration  constant and 
\bea
\ti{\ga}_{+}= & &\ga_{+}-\beta_{+}\fr{\si_{+}}{\de_{+}}\nonumber \\
            = & &
             \fr{2K_{4}\om_{3}}{K_{2}}\left (1+\fr{m}{M}\right )
             \left( \fr{K_{3}^{2}}{K_2K_4}-\fr{3}{4}\right ).
\eea
Eq.\,(38) is standard nonlinear Schr\"{o}dinger(NLS) equation.
It has an uniform vibrating solution 
\bea
\ti{f}_{+}=f_{0}\exp (i\ti{\ga}_{+}|f_{0}|^{2}t),\nonumber
\eea
where $f_{0}$ is any
complex constant, which corresponds to the linear optical upper cutoff 
phonon mode with a simply frequency shift $\ti{\ga}_{+}|f_{0}|^{2}$
and is a fixed point of the system. Note that it is possible to
eliminate the time dependence by a simply transformation, justfying
our use of the term ``fixed point'' for the uniform vibrating
solution. In fact, the fixed point may also be written as
\bea
\ti{f}_{+}=f_{0}\exp [i(\ti{\ga}_{+}f_{0}^{2}t+\bar{\phi})],\nonumber
\eea
where $f_0$ in this case is any real constant and $0\leq \bar{\phi}<2\pi$.
In this sense there is a ring of fixed point characterized
by the different values of the phase, $\bar{\phi}$. It is easy to show
that, since 
$\al_{+}<0$, when $\ti{\ga}_{+}<0$ the fixed point is unstable by 
a long wavelength small perturbation. 
This kind of  instability is due to a sideband modulation of the linear 
optical upper cutoff mode. The modulational instability for  
waves is called the 
Benjamin-Feir(BF) instability\cite{ben}. It is similar to 
the Eckhaus  instability for patterns in extended dissipative systems out of 
equilibrium.\cite{eck} 
The BF instability in discrete lattices and related formation of solitonlike 
localized states were already discussed by Kivshar and Peyrard.\cite{kiv3}
By this mechanism\,(usually called the Benjamin-Feir resonance 
mechanism\cite{stu})  
 a  linear optical upper cutoff excitation will 
bifurcate, grow exponentially at first and then saturate 
due to the nonlinearity of the system. At later stage, a nonlinear 
localized mode\,---\, optical upper cutoff soliton is formed.
In fact, for $\ti{\ga}_{+}<0$, {\it e.\,g}.,
\be
\fr{K_{3}^{2}}{K_2K_4}<\fr{3}{4},
\ee
Eq.\,(38) admit the envelope(breather) soliton solution
\be
\ti{f}_{+}=\left(\fr{\al_{+}}{\ti{\ga}_{+}}\right)^{\fr{1}{2}}\eta_{0}
           {\rm sech}\,[\eta_0(x_n-x_{n_0})]\exp
           [-i\fr{1}{2}|\al_{+}|\eta_{0}^{2}t-i\phi_0],
\ee
where $\eta_0$, $\phi_0$ and $x_{n_0}=n_0d$ are constants, $n_0$ is an
arbitrary integer. 
Inequality (41) is just the condition of the modulational instability for the 
linear optical upper cutoff mode. From 
Eq.\,(39) we obtain
\be
g_+=-\fr{\si_{+}\al_{+}}{\de_{+}\ti{\ga}_{+}}\eta_0
    \tanh [\eta_0 (x_n-x_{n_0})],
\ee
where the integration constant $C_1$ has chosen as zero as it corresponds to
a constant displacement for all particles. In leading approximation  the
lattice configuration takes the form
\bea
&  &  v_{n}(t)=-\fr{\si_{+}\al_{+}}{\de_{+}\ti{\ga}_{+}}\eta_0
      \tanh [\eta_0 (n-n_0)d]\nonumber\\
&  &  \hspace{15mm}+2\left(\fr{\al_{+}}{\ti{\ga}_{+}}\right)^{\fr{1}{2}}\eta_{0}
           {\rm sech}\,[\eta_0(n-n_0)d]\cos
           (\Omega_{3s}t+\phi_0),\\
&  &  w_{n}(t)=-\fr{\si_{+}\al_{+}}{\de_{+}\ti{\ga}_{+}}\eta_0
            \tanh [\eta_0 (n-n_0)d]\nonumber\\
&  &   \hspace{15mm}-2\fr{m}{M}\left(\fr{\al_{+}}{\ti{\ga}_{+}}\right)^{\fr{1}{2}}\eta_{0}
           {\rm sech}\,[\eta_0(n-n_0)d]\cos
           (\Omega_{3s}t+\phi_0),
\eea
with 
\be
\Omega_{3s}=\om_3+\fr{1}{2}|\al_{+}|\eta_{0}^{2},
\ee
{\it i.\,e}. the vibrating frequency of the localized mode is above the spectrum
of the linear optical mode thus above the all phonon bands. Hence (44) and (45)
represent  an ILM  accompanied by  an  asymmetric dc displacement 
due to the cubic anharmonicity
of the system. We call it the small-amplitude 
{\it asymmetric intrinsic localized mode}. 

From (44) and (45) we can see that the free parameter $\eta_0$ can be
taken as an expansion parameter, {\it i.\,e}. $\eta_0=O(\ep)$. 
By (46) we have $\eta_0=[2(\Omega_{3s}-\omega_3)/|\al_+|]^{1/2}$.
Thus in our approach, the expansion parameter $\ep$, used in (8),
is propotional to the squareroot of frequency difference between
the nonlinear localized mode and the linear cutoff phonon mode.

When $\ti{\ga}_{+}>0$, {\it e.\,g}., the inequality (41) takes opposite sign,
the uniform vibrating solution of the NLS equation (38) is neutral stable. In this case
Eq.\,(38) admits the dark soliton solution
\be
\ti{f}_{+}=\left(\fr{|\al_{+}|}{\ti{\ga}_{+}}\right)^{\fr{1}{2}}\eta_{0}
           \tanh [\eta_0(x_n-x_{n_0})]\exp
           [i|\al_{+}|\eta_{0}^{2}t-i\phi_0].
\ee
From  Eq.\,(39) we can obtain $g_{+}$ by integration. In this case we chose
$C_1$ in such a way\cite{fly1} that $(\pa g_{+}/\pa x_n)|_{|x_n|=\infty}=0$. Then we have 
$C_1=\si_{+}|\al_{+}|\eta_{0}^{2}/(\de_{+}\ti{\ga}_{+})$. Hence we have 
\be
g_+=\fr{\si_{+}|\al_{+}|}{\de_{+}\ti{\ga}_{+}}\eta_0
    \tanh [\eta_0 (x_n-x_{n_0})].
\ee
The lattice displacement in this case takes the form
\bea
&  &  v_{n}(t)=\fr{\si_{+}|\al_{+}|}{\de_{+}\ti{\ga}_{+}}\eta_0
      \tanh [\eta_0 (n-n_0)d]\nonumber\\
&  &  \hspace{15mm}+2\left(\fr{|\al_{+}|}{\ti{\ga}_{+}}\right)^{\fr{1}{2}}\eta_{0}
           \tanh [\eta_0(n-n_0)d]\cos
           (\Omega_{3k}t+\phi_0),\\
&  &  w_{n}(t)=\fr{\si_{+}|\al_{+}|}{\de_{+}\ti{\ga}_{+}}\eta_0
            \tanh [\eta_0 (n-n_0)d]\nonumber\\
&  &   \hspace{15mm}-2\fr{m}{M}\left(\fr{|\al_{+}|}{\ti{\ga}_{+}}\right)^{\fr{1}{2}}\eta_{0}
           \tanh [\eta_0(n-n_0)d]\cos
           (\Omega_{3k}t+\phi_0)
\eea
with 
\be
 \Omega_{3k}=\om_3-\fr{|\al_{+}|\ga_{+}}{\ti{\ga}_{+}}\eta_{0}^{2}.
\ee
Since $\ga_{+}<0$, the vibrating frequency of the kink mode
denoted by the expressions (49) and (50) is {\it greater} than $\om_3$. 
This is an example of a kink  with the vibrating frequency 
above the all phonon bands
due to the cubic nonlinearity of the system. 

\vspace{2mm}

(2). {\bf Optical lower cutoff mode}.  For the optical mode at 
$q=\pi/d$\,(zone-boundary optical phonon mode), one  has 
$\om_{+}=\om_2=\sqrt{2K_2/m}$,
$V_{g}^{+}=0$, $x_{n}^{+}=x_n$,
$\al_{+}=K_{2}d^{2}/[2\om_{2}(M-m)]$, 
$\beta_{+}=-K_{3}\om_{2}d/(2K_{2})$, 
$\ga_{+}=-3K_{4}\om_{2}/(2K_2)$, 
$\de_{+}=K_{2}^{2}d^{2}/(Mm)$ 
and $\si_{+}=4K_{2}K_{3}d/(Mm)$. 
In this case we have
\be
\tilde{\ga}_{+}=
\fr{2K_4}{K_2}\om_2\left (\fr{K_{3}^{2}}{K_2K_4}-\fr{3}{4}\right ).
\ee
If $\tilde{\ga}_{+}>0$, {\it i.\,e}.
\be 
\fr{K_{3}^{2}}{K_{2}K_{4}}>\fr{3}{4},
\label{cri2}
\ee
Eqs.\,(32) and (33) for plus sign have the solution
\bea
&  &  f_{+}=\left(\fr{\al_{+}}{\ti{\ga}_{+}}\right)^{\fr{1}{2}}\eta_{0}
           {\rm sech}\,[\eta_0(x_n-x_{n_0})]\exp
           [i\fr{1}{2}\al_{+}\eta_{0}^{2}t-i\phi_0],\\
&  &  g_+=-\fr{\si_{+}\al_{+}}{\de_{+}\ti{\ga}_{+}}\eta_0
    \tanh [\eta_0 (x_n-x_{n_0})].
\eea 
The lattice configuration takes the form
\bea
&  &  v_{n}(t)= -\fr{\si_{+}\al_{+}}{\de_{+}\ti{\ga}_{+}}\eta_0
      \tanh [\eta_0 (n-n_0)d]\nonumber\\
&  &  \hspace{15mm}+(-1)^{n}2\left(\fr{\al_{+}}{\ti{\ga}_{+}}\right)^{\fr{1}{2}}\eta_{0}
           {\rm sech}\,[\eta_0(n-n_0)d]\cos
           (\Omega_{2s}t+\phi_0),\\
&  &  w_{n}(t)= -\fr{\si_{+}\al_{+}}{\de_{+}\ti{\ga}_{+}}\eta_0
            \tanh [\eta_0 (n-n_0)d]
\eea
with 
\be
\Omega_{2s}=\om_2-\fr{1}{2}\al_{+}\eta_{0}^{2},
\ee
lying in the frequency gap of the phonon spectra 
between the acoustic and the optical modes.
It is a typical asymmetric nonlinear gap mode,  existing in the diatomic lattices
when the condition (53) is satisfied. We note that for this mode 
the displacement of the heavy particles only has a kink-like asymmetric dc  
part. But the displacement of
the light particles,  besides  the same  type of  dc part,
has an additional ``staggered'' vibrational part\,({\it i.\,e}. 
``staggered'' envelope soliton).  
We call it the {\it asymmetric optical lower cutoff gap soliton}. 
The vibrating frequency $\Omega_{2s}$ has the parabola relation with respect
to
the wave amplitude, denoted by the parameter $\eta_0$. The formation of the
asymmetric nonlinear gap mode is also the conclusion of the BF instability
for the corresponding linear optical lower cutoff phonon mode. A further 
discussion for such nonlinear modes in the realistic potentials is
given in the next section.

\vspace{2mm}

(3). {\bf Acoustic upper cutoff mode}. For the acoustic mode at 
$q=\pi/d$\,(zone-boundary acoustic phonon mode), one  has 
$\om_{-}=\om_1=\sqrt{2K_2/M}$,
$V_{g}^{-}=0$, $x_{n}^{-}=x_n$,
$\al_{-}=-K_{2}d^{2}/[2\om_{1}(M-m)]<0$, 
$\beta_{-}=-K_{3}\om_{1}d/(2K_{2})$, 
$\ga_{-}=-3K_{4}\om_{1}/(2K_2)$, 
$\de_{-}=K_{2}^{2}d^{2}/(Mm)$
and  
$\si_{-}=4K_{2}K_{3}d/(Mm)$. 
Similarly
one can obtain  the equations like (38) and (39) with
$g_{+}, \tilde{f_{+}}, \si_{+}, \de_{+}, \al_{+}$ and $\tilde{\ga}_{+}$  
changed by
$g_{-}, \tilde{f_{-}}, \si_{-}, \de_{-}, \al_{-}$ and $\tilde{\ga}_{-}$. 
Here
\be
\tilde{\ga}_{-}=\ga_{-}-\beta_{-}\fr{\si_{-}}{\de_{-}}=
\fr{2K_4}{K_2}\om_1\left (\fr{K_{3}^{2}}{K_2K_4}-\fr{3}{4}\right ).
\ee
If $\tilde{\ga}_{-}<0$, when $\tilde{\ga}_{-}<0$, {\it i.\,e}.
\be 
\fr{K_{3}^{2}}{K_{2}K_{4}}<\fr{3}{4},
\label{cri3}
\ee
due to a BF instability of the corresponding linear upper cutoff acoustic
mode
an {\it asymmetric acoustic upper cutoff gap soliton} appears, with the vibrating 
frequency being in the gap of the phonon spectra between the acoustic and
optical modes. Otherwise an acoustic 
upper cutoff kink vibrational mode occurs. 
We can readily written down the explicit expression of 
the lattice configuration in this case, but for
saving space it is omitted here. 

\vspace{2mm}

(4). {\bf Acoustic lower  cutoff mode}. This is a long wavelength mode
without any carrier wave,  because when $q=0$ we have $\om_{-}=0$ thus 
$\phi_{n}^{-}=0$. The lattice displacement only has dc part and its evolution
is controled by the MMKdV equation,  given by (34). A single-soliton solution
of Eq.\,(34) is given by 
\be 
u=\fr{24\kappa^{2}H}
{P+\sqrt{P^2+24\kappa^2 QH}\cosh\,[2\kappa (nd-(c+24\kappa^2 H)t)]},
\ee
where $\kappa$ is an arbitrary constant. Making the transformation
$u=-P/(2Q)+U(\zeta_{n},t)$ with 
$\zeta_{n}=x_n+P^2 t/(4Q)$, Eq.\,(34) becomes
\be
\fr{\pa U}{\pa t}+Q\,U^2\fr{\pa U}{\pa \zeta_n}+H\,\fr{\pa^3 U}{\pa \zeta_{n}^{3}}=0.
\ee
It  is the  standard MKdV equation and can be solved by
the inverse scattering transform. For the explict expressions of 
the kink and breather solutions
of the MKdV equation we refer to Ref.\,11.

Needless to say, in addition to the cutoff modes considered above, our approach
developed in Sec.\,II can also be used to discuss the nonlinear localized
excitations for $q\neq 0$ and $q\neq \pi/d$\,({\it i.\,e}. the intra-band modes),
which will be done elsewhere.
\section{APPLICATION TO REALISTIC TWO-BODY NEAREST-NEIGHBOR POTENTIALS}
In this section,  we apply the general results obtained above to
the diatomic lattices with the standard two-body nearest-neighbor potentials
to see whether the anharmonic gap modes and the ILM's can appear or not.
This can be easily done by using the existence criteria given by 
(41)\,(for ILM's), (\ref{cri2})\,(for optical lower cutoff gap solitons), and 
(\ref{cri3})\,(for acoustic upper cutoff gap solitons). Four standard interatomic
potentials are:\cite{kis3} 
\begin{enumerate}
\item {\it Toda}:
 \be
 V(r)=\fr{a}{b}{\rm e}^{-br}+ar-\fr{a}{b},
 \ee
 where $a$ and $b$ are coefficients, such that $ab>0$. $r$ is the deviation
 of the relative interparticle separation from its zero-temperature
 equilibrium position.

\item {\it Born-Mayer-Coulomb}:
 \be
 V(r)=\fr{\al_{M}q^2}{d_0^2}\left [-\fr{d_0^2}{r+d_0}
      +\rho_0 {\rm e}^{-r/\rho_0}+d_0-\rho_0\right ],
 \ee
 where $\al_{M}$ is the Madelung constant, $q$ is the effective charge, 
 $d_0$ is the zero-temperature equilibrium
 distance between adjacent particles, and $\rho_0$ is the 
 constant describing the repulsion between atoms.
\item {\it Lennard-Jones}:
 \be
 V(r)=\al \left [\left(\fr{d_0}{r+d_0}\right )^{12}
      -2\left (\fr{d_0}{r+d_0}\right )^{6}+1\right ],
 \ee
 where $\al$ is the constant determing the potential strength.
\item {\it Morse}:
 \be
 V(r)=P\left ({\rm e}^{-ar}-1\right )^{2},
 \ee
 where $P$ and $a$ are constants determing the strength and the curvature of
 the potential, respectively.
\end{enumerate}

For weakly nonlinear excitations we can Taylor expand these potentials 
at their equilibrium position\,($r=0$) to obtain
the force constants. They are defined by
\be
K_{j}=\fr{1}{(j-1)!}\left (\fr{{\rm d}^{j}V}{{\rm d}r^j}\right )_{r=0}
\ee
with $j=2, 3, 4,\cdots$. 
Thus we have the values of $K_2$, $K_3$,  
$K_4$ and $K_3^2/(K_2K_4)$ which are given in Table 1.
Notice that $K_3<0$ for these potentials except that 
the Toda one with $a$ and $b$ being both negative\,(note that we require that
$K_2$ and $K_4$ are positive).
The parameter $I$ for the Born-Mayer-Coulomb potential in Table 1 is
defined by
\be 
I=\fr{(d_0^2-6\rho_0^2)^2}{(d_0-2\rho_0)(d_0^3-24\rho_0^3)}.
\ee
Generally,  we have  $I>1$. For example\cite{san,kis3,bor},
for KI, $d_0$=3.14\AA, $\rho_0=0.26$\AA, one has $I=1.1171$;
for KBr, $d_0=3.29$\AA,  $\rho_0=0.334$\AA, we have $I=1.1328$; 
for LiI, $d_0=3.0$\AA, $\rho_0=0.374$\AA, one has $I=1.1487$.
By these  results  as well as the criteria given by  (41), (53) and
(60), we make the following conclusions:

\vspace{2mm}
(1). The condition (53) is satisfied by all these standard two-body
potentials. Thus the asymmetric optical lower cutoff gap soliton
mode given in the expressions (56) and (57) {\it does} exist in the
diatomic lattices with the Toda, Born-Mayer-Coulomb, 
Lennard-Jones and Morse type interatomic interactions.  This result agrees
with the conclusion by numerical study presented by Kiselev 
{\it et al}.\cite{kis2,kis3}

Shown in Fig.\,1 is the dc\,(denoted by the solid circles) and
ac\,(denoted by the open circles) amplitude patterns of light
particles for the optical lower cutoff gap soliton mode. 
The amplitude pattern for the heavy particles, which
is not shown here,  only has dc
part similar to that of the dc part for the light particles.
The Born-Mayer-Coulomb potential for KBr-like parameters is used
with $m/M=39/80$. Comparing the panel (a) of Fig.\,1 in Ref.\,34
with our result shown in Fig.\,1 here, we see that the lattice 
configuration obtained analytically by our  QDA is basically the
same as the corresponding result given by the rotating wave
approximation plus computer simulation, used by
Kiselev {\it et al}.\cite{kis3}  
Furthermore, the parabola relation
between the vibrating frequency and the amplitude of the nonlinear
optical lower cutoff gap mode, given by (58), is in accordance  well with
the numerical results\,(see the Fig.\,3 of Ref.\,34).

Shown in Fig.\,2 is the amplitude\,(maximum absolute value) 
ratio $(|A_{dc}|/|A_{ac}|)$ of the dc
         part\,($|A_{dc}|$) to ac part\,($|A_{ac}|$)
of the light particle displacement for the optical lower cutoff gap 
soliton mode. We see that the amplitude ratio is the function of the mass
ratio $m/M$. The larger is $m/M$, the larger is the amplitude ratio.
The amplitude ratio grows very  fast as the mass ratio approaches 1.

\vspace{2mm}  
(2). Notice that the existence criterion of an ILM\,(the optical
upper cutoff soliton) is the inequality (41), {\it i.\,e}.
the necessary condition for the occurence of the ILM is
\be
\fr{K_{3}^{2}}{K_2K_4}<\fr{3}{4}.
\ee
From Table 1 we see that in general $K_3^2/(K_2K_4)>1$. Thus for all the
standard two-body potentials from Toda to Morse type,  {\it an ILM 
is impossible}. This conclusion concides also with the numerical
results of Refs.\,33 and 34.

\vspace{2mm}
(3). An acoustic upper cutoff gap soliton is not possible for all
these standard two-body potentials, because its  existence condition
(60)\,(the same as (41)\,) can not be satisfied. However, 
acoutic and optical upper cutoff
vibrating kinks are possible excitations for these diatomic lattice
systems.

\vspace{2mm}
(4). Nonlinear acoustic lower cutoff excitations are long wavelength modes
and are governed by the MMKdV equation (34). Since $Q$ and $H$ are positive
and $P$ is negative\,(due to $K_3<0$)
 for all the two-body potentials\,(except the Toda one with 
 $a$ and $b$ being both negative), the soliton 
amplitude\,(see (61)\,) in the presence of cubic nonlinearity ($K_3\neq0$)
is smaller than the soliton amplitude in the absence of the cubic 
nonlinearity($K_3=0$).

\vspace{2mm}
(5). For general $K_2$-$K_3$-$K_4$ potentials, by the criteria given by
(41), (53) and (60), we conclude that  in weakly nonlinear approximation
the ``critical'' value of $K_3^2/(K_2K_4)$ for the transition from an
optical upper cutoff kink to an optical upper cutoff soliton\,(ILM) 
and for the
occurence of an optical(acoustic) lower(upper) cutoff gap mode
is 3/4, independent of $m/M$. Because one of the conditions (53) and (60)
must be satisfied, we have the conclusion that {\it for any nonlinear
diatomic lattice, gap solitons always occur}. Our analytical results
support the numerical findings of Franchini {\it et al}.\cite{fra}: 
for small cubic nonlinearity, nonlinear optical and acoustic 
upper cutoff localized modes appear, while for large cubic nonlinearity
a nonlinear optical lower cutoff mode rises. When $K_3=0$, the corresponding
theoretical and numerical results for harmonic plus quartic 
potentials\cite{hua2,fra} are recovered. 

It is interesting to note that 
the criterion  (41) and (60) for the occurence of the 
asymmetric optical and acoustic upper cutoff solitons 
in the diatomic lattices are  the same as that for the appearence
of the upper cutoff solitons in monatomic lattices,
given by Tsurui\cite{tsu}\,(Eq.\,(4.8), $\om_c^2=4$),
Flytzanis {\it et al}.\cite{fly1}\,(Eq.\,(5.5), $kD=\pi$) and
Flach\cite{fla3}\,(Eq.\,(3.24), $v_4=0$). This criterion was also
given implicitly in Ref.\,10 since the envelope soliton solution
(23) in Ref.\,10 requires ${\rm sgn}(PQ)>0$.

In addition, our results show analytically that for the potentials with cubic
and quartic  nonlinearity  
an nonlinear localized excitation always consists of dc and ac 
parts. The appearrence of the 
dc part is a direct  conclusion of the asymmetry in the 
potentials. This fact is already known by using different 
approaches\,(see {\it e.\,g}. Refs.\,33 and 34).

Recently, Bonart, R\"{o}ssler and Page\cite{bon1} considered 
the boundary condition effects in the diatomic lattice
with cubic and quartic anharmonicity and gave existence
criteria for the ILM's and the IGM's 
not restricted to small amplitudes by using a 
different approach. By comparison
we can see that our instability threshold criterion
for the linear optical upper cutoff phonon modes is in accordance with
their  corresponding 
one\,[Eq.\,(10) in Ref.\,45)] when exact to $A_{ac}^2$\,[remenber 
that in our notation $A_{ac}$ is the amplitude of the
ac part of the nonlinear excitation, {\it e.\,g}. in (44) we have 
$A_{ac}=2(\al_{+}/\tilde{\ga}_{+})^{1/2}\eta_0$]. It is easy to
show that our frequency formulas for the nonlinear optical 
(including upper and lower)  cutoff modes, {\it i.\,e}. Eqs.\,(46) 
and (58), also coincide with the corresponding
ones in Ref.\,45\,[Eqs.\,(9) and (12)] if the terms propotional to
$A_{ac}^4$ are neglected. Our results support also the criterion 
(14) in Ref.\,45 for the appearance of nonlinear localized modes.
In addition, our analytical approach based on QDA may provide much more 
insights  such as for the occurance of nonlinear localized 
excitations in acoustic branch  although it has the limitation of small
amplitude.
\section{DISCUSSION AND SUMMARY}
Based on the QDA for diatomic lattices we have {\it analytically} studied 
the nonlinear localized excitations with small amplitude
in the diatomic lattices with cubic
and quartic nonlinearity. The results are quite general
and allows us directly to obtain many different types of
nonlinear excitations in a unified way.  Starting from the nonlinear 
amplitude equations given in (32) and (33),
the existence criteria for the optical upper cutoff solitons 
and asymmetric gap soliton modes 
have been  explicitly provided in (41), (53) and (60). The analytical
expressions of particle displacements for all these nonlinear 
localized modes are also given.
The theoretical results have been 
applied to the standard two-body nearest-neighbor potentials
from Toda to Morse type and agreements well with the previous numerical 
findings have also been found.

Most of the existed analytical studies for the nonlinear excitations
in diatomic lattices involved so-called ``decoupling ansatz'', in which
some relations were assumed between the displacement of light particles
and that of heavy ones before solving the equations of motion\,(see
Ref.\,11 and references therein). Pnevmatikos {\it et al}.\cite{pne2}
investigated the soliton dynamics of nonlinear diatomic lattices by
using the decoupling ansatz. For long wavelength\,({\it i.\,e}. $q=0$)
excitations this can be done without big difficulty. But for envelope-type
excitations and in the case of cubic nonlinearity,  the concrete form of
the decoupling ansatz is not easy to determine and the 
analytical calculation
for the coefficients in amplitude equations, like (32) and (33),
is also heavy. Except  several
numerical studies,  such an analytical calculation has never been accomplished.
Recently, it was shown that the decoupling ansatz is completely
unnecessary and can be derived by the QDA\cite{hua2}. In addition, the QDA
has many other advantages. For example, the results obtained by the QDA,
though restricted to small amplitudes,
are valid in the whole BZ except at the zero-dispersion point of
the optical phonon branch\,( see Ref.\,46). 
Thus one can obtain all solutions for nonlinear
cutoff and non-cutoff modes in a simple and unified way; the method is
quite general and can be applied to the other lattice systems. An extension
based on the QDA for  magnetic gap soliton excitations
 in alternating Heisenberg 
ferromagnets has been given recently.\cite{hua5}  
In the present work, we use the QDA to consider the nonlinear localized
excitations in the diatomic lattices with cubic and quartic nonlinearity.
The detailed expressions of the coefficients in the amplitude equations
(32) and (33) as well as (34) thus various types of nonlinear excitations
are {\it explicitly} given.

The study of stability by using nonlinear amplitude equations is widely 
employed in pattern formation in systems out of equilibrium.\cite{stu,cro}
In present work, by a similar idea we have studied the stability of the
linear optical and acoustic (upper and lower) cutoff phonon modes.
For the $K_2$-$K_3$-$K_4$ potential, we have obtained the existence criteria
for the occurance of linear cutoff phonon mode-related  nonlinear
localized excitations: If $K_{3}^2/(K_2K_4)<3/4$,  we  have the acoustic 
and optical upper cutoff solitons. Otherwise one has only the 
optical lower cutoff (gap) solitons. The later case happens
for the standard two-body potentials from Toda to Morse type.  
The formation of the asymmetric gap solitons in the standard two-body
potentials is the result of BF modulational instability of the optical
lower cutoff phonon modes. Our results show that for any nonlinear diatomic
lattice a gap soliton always occurs. The reason for this is that
the curvatures of the acoustic and optical branches of the phonon
spectra in the vicinity of the BZ edge have different signs. This means
that if the optical mode at the BZ edge supports a 
lower cutoff gap soliton, the acoustic mode at the BZ edge supports an upper
cutoff kink subject to the same sign of the nonlinearity, and vice vesa.

For large-amplitude excitations we should extend the present QDA to
a higher-order approximation. It is expected that higher-order 
correction terms, likely 
$|f_{\pm}|^{4}f_{\pm}$ and $(\pa/\pa x_n^{\pm})|f_{\pm}|^4$\,(as well
as some higher-order dispersion terms), 
will be added respectively into Eqs.\,(32) and (33). Then 
the existence criteria for the nonlinear localized excitations,
{\it i.\,e}. (41), (53) and (60), will be modified to be 
amplitude-dependent.

The generation of an
asymmetric IGM in nonlinear 3D diatomic lattices
has been considered recently\cite{bon,kis4}. These studies reveal the 
possibility
of observing experimentally the IGM's in real (3D) crystals. An macro-analogy  
of the diatomic lattices with cubic and quartic nonlinearity is to use
a chain of magnetic pendulums. Russell {\it et al}.\cite{rus} reported 
moving breathers in such a system quite recently,
but they did not pay attention to nonlinear
cutoff modes. If the mass of the magnetic pendulums is arranged in a 
alternating way, one can obtain a diatomic lattice with an asymmetric 
intersite\,(dipole-dipole interaction) potential. For small amplitude 
excitations the
intersite potential reduces to a $K_2$-$K_3$-$K_4$ one. Thus it is possible
to observe the asymmetric IGM's by using this macro-diatomic lattice
system. 
\section*{ACKNOWDGEMENTS}
G.\,H.  wishes to express his  appreciation to Prof. Director P. Fulde for the warm 
hospitality received at the Max-Planck-Institut f\"{u}r Physik komplexer
Systeme, where part of this work was carried out.  
It is a pleasure to thank Prof. A. J. Sievers for kindly sending a copy
of Ref.\,21, Dr. S. Flach for the critical reading of the manuscript, 
Drs. M. B\"{a}r and A. Cohen,  Profs. F. G. Mertens
and M. G. Velarde for fruitful discussions. We also thank an 
anonymous referee for bringing our attention to Ref.\,45 
and useful comments.
This work was partly supported by the Hong Kong Research Grant Council
and the Hong Kong Baptist University Faculty Research Grant. 
\newpage
\section*{APPENDIX A}
The detailed expressions of the coefficients of Eqs.\,(22) and (23) are 
given by
$$\displaylines{
\de_{-}=I_2J_2d^2-2(I_2+J_2)\left[\fr{I_2J_2d\sin (qd)}
        {2\om_{-}(I_2+J_2-\om_{-}^2)}\right ]^2,
        \hfill \rlap{(A.1)}\cr
\si_{-}=4I_2I_3d\fr{\om_{-}^2-2J_2}{\om_{-}^2-2I_2}+
        \fr{4J_2I_3d}{\om_{-}^2-2I_2}(\om_{-}^2+2I_2\cos qd)
        +\fr{32I_2J_2I_3V_{g}^{-}\om_{-}\sin qd}
         {(\om_{-}^2-2I_2)^2},
         \hfill \rlap{(A.2)}\cr
\al_{-}=\fr{4(I_2+J_2-3\om_{-}^2)(V_{g}^{-})^2
         -d^2[(\om_{-}^2-2I_2)(\om_{-}^2-2J_2)-2I_2J_2]}
         {4\om_{-}(\om_{-}^2-I_2-J_2)},
         \hfill \rlap{(A.3)}\cr
\beta_{-}=-\fr{I_3\om_{-}d}{2I_2},
         \hfill \rlap{(A.4)}\cr
\ga_{-}=-\fr{3J_4\om_{-}}{2J_2}
        \left [1+\fr{\om_{-}^2-2J_2}{J_2}\left(1+\fr{I_2}{\om_{-}^2-2I_2}\right )
        \right ]
        \hfill\cr
   \qquad+\fr{\om_{-}^3º\sin^{2}(qd)}{(\om_{-}^2-2I_2)(I_2+J_2-\om_{-}^2)}
        º\left[ -I_3L_{-}+\fr{J_3(I_3+I_2L_{-}\cos qd)}
        {2\om_{-}^{2}-I_2}\right ],
        \hfill \rlap{(A.5)}\cr
L_{-}=\fr{I_2J_3[\cos qd-(2\om_{-}^2-I_2)/J_2]}
      {(2\om_{-}^{2}-J_2)(2\om_{-}^2-I_2)-I_2J_2\cos^2qd}.
      \hfill \rlap{(A.6)}\cr
}$$
The coefficients of Eqs.\,(28) and (29) are given by
$$\displaylines{
\de_{+}=I_2J_2d^2-2(I_2+J_2)\left[\fr{I_2J_2d\sin (qd)}
        {2\om_{+}(I_2+J_2-\om_{+}^2)}\right ]^2,
        \hfill \rlap{(A.7)}\cr
\si_{+}=4J_2J_3d\fr{\om_{+}^2-2I_2}{\om_{+}^2-2J_2}+
        \fr{4I_2J_3d}{\om_{+}^2-2J_2}(\om_{+}^2+2J_2\cos qd)
        +\fr{32I_2J_2J_3V_{g}^{+}\om_{+}\sin qd}
         {(\om_{+}^2-2J_2)^2},
         \hfill \rlap{(A.8)}\cr
\al_{+}=\fr{4(I_2+J_2-3\om_{+}^2)(V_{g}^{+})^2
         -d^2[(\om_{+}^2-2J_2)(\om_{+}^2-2I_2)-2I_2J_2]}
         {4\om_{+}(\om_{+}^2-I_2-J_2)},
         \hfill \rlap{(A.9)}\cr
\beta_{+}=-\fr{J_3\om_{+}d}{2J_2},
         \hfill \rlap{(A.10)}\cr
\ga_{+}=-\fr{3I_4\om_{+}}{2I_2}
        \left [1+\fr{\om_{+}^2-2I_2}{I_2}\left(1+\fr{J_2}{\om_{+}^2-2J_2}\right )
        \right ]
        \hfill\cr
   \qquad+\fr{\om_{+}^3º\sin^{2}(qd)}{(\om_{+}^2-2J_2)(I_2+J_2-\om_{+}^2)}
        º\left[ -J_3L_{+}+\fr{I_3(J_3+J_2L_{+}\cos qd)}
        {2\om_{+}^{2}-J_2}\right ],
        \hfill \rlap{(A.11)}\cr
L_{+}=\fr{J_2I_3[\cos qd-(2\om_{+}^2-J_2)/I_2]}
      {(2\om_{+}^{2}-I_2)(2\om_{+}^2-J_2)-I_2J_2\cos^2qd},
      \hfill \rlap{(A.12)}\cr
}$$
where $\om_{\pm}(q)$ has been given in Eq.\,(7).

\newpage

\vspace*{3cm}

\large

\vspace{4cm}
\begin{table}
 \caption{
The force constants $K_2$, $K_3$, $K_4$
                      and the value of $K_3^2/(K_2K_4)$ for the standard
        two-body potentials from the Toda,
         Born-Mayer-Coulomb\,(B-M-C),
        Lennard-Jones\,(L-J), and Morse type.
            }

 \begin{center}
   \begin{tabular}{c c c c c }
   potential  &$K_2$  &$K_3$          &$K_4$         &$K_3^2/(K_2K_4)$\\
   \tableline
   Toda       &$ab$   &-$\fr{ab^2}{2}$ &$\fr{ab^3}{6}$ &$\fr{3}{2}$\\
   \tableline
   B-M-C      &$\fr{\al_Mq^2(d_0-2\rho_0)}{\rho_0d_0^3}$
              &-$\fr{\al_Mq^2(d_0^2-6\rho_0^2)}{2\rho_0^2d_0^4}$
              &$\fr{\al_Mq^2(d_0^3-24\rho_0^3)}{6\rho_0^3d_0^5}$
              &$\fr{3}{2}I$\\
   \tableline
   L-J        &$\fr{72\al}{d_0^2}$
              &$-\fr{756\al}{d_0^3}$
              &$\fr{6678\al}{d_0^4}$
              &$\fr{63}{53}$\\
   \tableline
   Morse      &$2Pa^2$
              &$-Pa^3$
              &$\fr{Pa^4}{3}$
              &$\fr{3}{2}$
    \end{tabular}
 \end{center}
\end{table}

\begin{figure}
\centerline{\psfig{figure=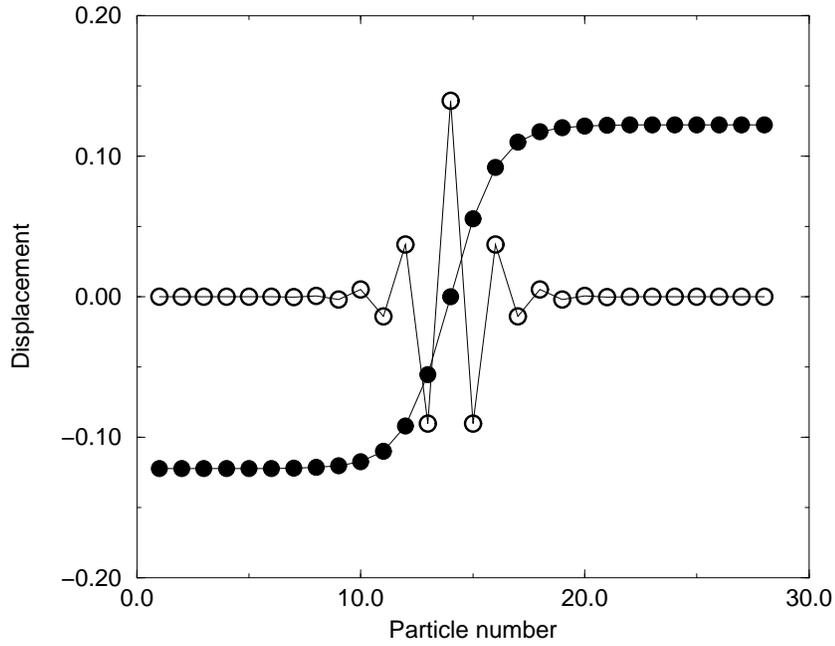,height=10cm,angle=0}}
\caption{The dc and ac amplitude patterns of light 
         particles in a diatomic lattice for the 
        optical lower cutoff gap soliton mode. The
         solid\,(open) circles denote dc\,(ac) part
        of the lattice displacement. The Born-Mayer-Coulomb 
        potential for KBr-like 
        parameters is used  with $m/M=39/80$.}
\label{fig:fig1}
\end{figure}

\begin{figure}
\centerline{\psfig{figure=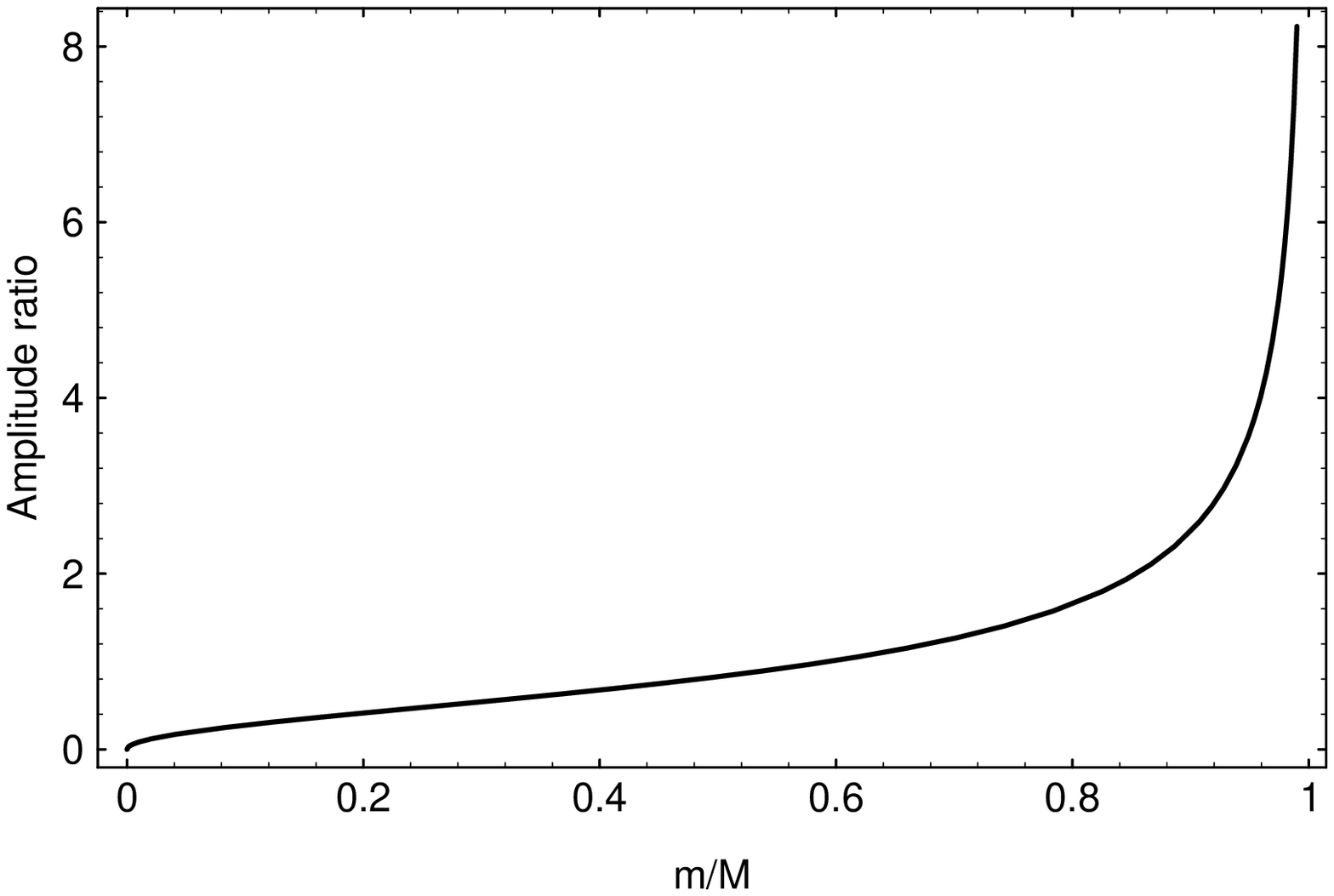,height=15cm,angle=0}}
\caption{
The amplitude(maximum absolute value) ratio 
         $(|A_{dc}|/|A_{ac}|)$ of the dc
         part\,($|A_{dc}|$)
       to ac part\,($|A_{ac}|$) of the light particle
         displacement for the optical lower cutoff gap
         soliton mode
        as a function of the mass ratio  $m/M$.
          }
\label{fig:fig2}
\end{figure}


\begin{references}
\bibitem[46]{for}For the optical phonon branch a wavenumber $q=q_c$
                 exists at which the curvature of $\omega^{+}(q)$, 
                 $2\al_{+}={\rm d}^2\omega_{+}/{\rm d}q^2$, 
                 vanishes\,({\it i. e}. $q=q_c$ is a zero-dispersion point). 
                 Thus at $q=q_c$ the Eq.\,(28) is not 
                 valid for the description of nonlinear excitations.
                 In this case a higher-order dispersion term
                 $-\fr{i}{3!}{\rm d}^3\omega^{+}/{\rm d}q^3 G_{11}$
                 should be added in the Eq.\,(28).
\end{references}
\end{document}